\documentclass[10pt]{iopart}
\usepackage{amscd,amssymb}
\usepackage{graphicx}
\newcommand{\eqref}[1]{(\ref{#1})}

\newcommand{\diff}{\mathrm{d}}

\begin{document}
\title{Ground-state properties of a supersymmetric fermion chain}
\author{Paul Fendley$^{1,3}$ and Christian Hagendorf\,$^{2,3}$}
\address{$^1$ Microsoft Station Q, University of California, Santa Barbara, CA 93106}
\address{$^2$ Kavli Institute for Theoretical Physics, University of California, Santa Barbara, CA 93106}
\address{$^3$ Department of Physics, University of Virginia, 382
  McCormick Road, Charlottesville, VA 22904}

\ead{fendley@virginia.edu, hagendorf@virginia.edu}

\begin{abstract}
We analyze the ground state of a strongly interacting fermion chain
with a supersymmetry. We conjecture a number of exact results, such as a
hidden duality between weak and strong couplings. By exploiting a
scale free property of the perturbative expansions, we find exact
expressions for the order parameters, yielding the
critical exponents. We show that the ground state of this fermion
chain and another model in the same universality class, the XYZ chain
along a line of couplings, are both written in terms of the same
polynomials.  We demonstrate this explicitly for up to $N=24$ sites, and
provide consistency checks for large $N$. These polynomials satisfy a
recursion relation related to the Painlev\'e VI differential
equation, and using a scale-free property of these polynomials, we
derive a simple and exact formula for their $N\to\infty$ limit.
\end{abstract}


\section{Introduction}

The calculation of exact results for physical quantities in
interacting one-dimensional quantum chains has a long and
distinguished history, going back to the early days of quantum
mechanics. Bethe's results for the Heisenberg chain more than 75 years
ago were the first; there are now a whole host of models that are
often referred to as ``solved''.

The word ``solved'' sometimes means that the exact ground state is
known in closed form, and it sometimes means that the model is
integrable. Integrability means that there exist enough conserved
charges to completely constrain the dynamics. Although it is often
possible to exploit the integrability to do exact computations,
usually it is not possible to find the ground state explicitly in an
integrable model. Conversely, many models where the ground state is
known explicitly are not integrable.

Even when we restrict our study to integrable models, there is still a
wide variation in what can be calculated and how difficult it is to do
so. On one end of the spectrum, there are systems that can be mapped
to free fermions, and so essentially everything can be computed
explicitly. The canonical example of this is the quantum Ising chain,
often referred to as the Ising model in a transverse field. On the
other end of the spectrum, there are models such as the XYZ chain,
where the standard Bethe ansatz does not work, and much more elaborate
methods are required to derive exact results \cite{Baxbook}.

It is our aim in this paper to discuss one example of a class of
integrable models that are seemingly difficult to solve, in that the
standard Bethe ansatz does not apply.  In fact, in the model discussed
in this paper, we have not even proved that it is integrable (although
we believe it is), much less exploited the integrability directly.
However, we will show that exact results can be found by using very
different methods than those usually utilized in integrability. 

An essential ingredient in our analysis is {\em supersymmetry}.  One
of the hallmarks of supersymmetric systems is the special properties
satisfied by the ground state \cite{Witten}. This usually makes it
possible to not only find the ground-state energy exactly and easily
for any number of sites $N$, but to find simple operators that
annihilate the ground state. A supersymmetric system need not be
integrable, but when it is, the ground state itself satisfies a number
of special properties, even for an integrable system. For example, we
will show here that this makes it possible to find exact values for
the critical exponents simply by analyzing the ground state at finite
$N$. It also makes it possible to numerically determine scaling
functions extremely accurately from very small system sizes.

The specific systems we are studying were introduced in
\cite{fendley:02}. These are chains of spinless fermions with strong
interactions where the supersymmetry is built in from the start, so
that the ground state of this many-body system has many special
properties. The supersymmetric fermion chains analyzed in depth
in \cite{fendley:02,fendley:03} are critical. Here we study a
one-parameter deformation of the critical supersymmetric chain.  As we
discussed briefly in an earlier paper \cite{fendley:10}, we find
simple but exact expressions for a variety of physical quantities. Here we extend
this work considerably, and find both scaling functions and exact results for the
ground state itself.

A famous example of a system in the class we are studying
is the XXZ chain at anisotropy $\Delta=-1/2$ (see e.g.\
\cite{razumov:01,stroganov:01_2,degier:02}). A host of elegant results
relating the ground state of this system to combinatorial quantities
have been found. One of the reason these properties were found is that
like with a supersymmetric system, the ground state energy is exactly
known. In fact, this XXZ chain has a hidden supersymmetry  \cite{fendley:03,yang:04}.
Our off-critical results on the fermion chain here
results are quite analogous to those found for an off-critical
deformation of the XXZ chain to the XYZ chain studied in
\cite{bazhanov:05,bazhanov:06,mangazeev:10,razumov:10,fendley:10}. In
fact, not only is this XYZ chain in the same
universality class as our fermion chain, but the ground states can be constructed in
terms of the same polynomials! These polynomials have a host of very
special properties that we analyze in depth.

We believe that supersymmetric integrable systems of this type  are ``more than
integrable'' -- the additional symmetries coming from supersymmetry
result in structure simply not present in an integrable system. In
fact, in some aspects, supersymmetric integrable systems are simpler
than free-fermion systems. For example, the ground state energy in a
free-fermi system such as the Ising chain depends non-trivially on the
size of the system: as one increases $N$, one must keep filling the
fermi sea. In our supersymmetric systems, the ground state energy is
exactly zero for any $N$. 

The layout of this paper is as follows. In section \ref{sec:model} we
introduce the lattice fermion model with supersymmetry.  Using both
the supersymmetry and an exact computation of the ground state up to
$N=24$, we discuss general and specific properties of the ground state
in section \ref{sec:gswf}. In particular, we show how the ground
states can be characterized by a polynomial with very special
properties. In section \ref{sec:dualityscaling}, we discuss the duality and
the scaling behavior of the order parameters. A connection to the XYZ
model, discussed in section \ref{sec:xyz}, indicates a relation of our
model with classically integrable equations. We exploit this to find a
recursion relation for the polynomials describing the ground state.
We then are able to find new results for these polynomials by
exploiting their scale-free behavior. We present our conclusions along
with some open questions in section \ref{sec:concl}.

\section{The model}
\label{sec:model}

In this section we introduce our model explicitly, and analyze some of
its basic properties. The degrees of freedom are ``fat'' spinless fermions on a periodic chain. They are
annihilated and created by operators $c_i$ and $c_i^\dagger$ obeying
the standard anticommutation relations
$\{c_i,c_j\}=\{c_i^\dagger,c_j^\dagger\}=0$ and
$\{c_i,c_j^\dagger\}=\delta_{ij}$.  What makes them fat is that,
following \cite{fendley:02}, we do not allow two adjacent sites to be
occupied. Defining the projector $P_j=1-c^\dagger_j c_j$, this means
we restrict the Hilbert space to include only configurations where
$P_jP_{j+1}=0$ for all $j$.  Denoting an empty site by $\circ$, and an
occupied site by $\bullet$, this hard-core constraint therefore
excludes pairs $\cdots \bullet\bullet \cdots$.

The reason we study fat fermions is that there exists a Hamiltonian
with ${\mathcal N}=2$ supersymmetry on this Hilbert space
\cite{fendley:02}. This is because the ``supercharges''
\begin{equation*}
  Q^- = \sum_{j=1}^N \lambda_j P_{j-1}c_j P_{j+1},\qquad Q^+ = \sum_{j=1}^N \lambda_j^\ast P_{j-1}c_j^\dagger P_{j+1}.
\end{equation*}
are both nilpotent: $(Q^-)^2=(Q^+)^2=0$ for any complex numbers
$\lambda_j$.  Here we take periodic boundary conditions on a chain of
length $N$, so all indices are interpreted mod $N$.
Defining the Hamiltonian to be the anticommutator
$H = \{Q^+,Q^-\}$, the nilpotency requires that both $Q^-$ and $Q^+$
commute with $H$. Moreover, the fermion anticommutators ensure that $H$
is local:
\begin{equation}
  H = \sum_{j=1}^N \left[P_{j-1}\left(\lambda_{j+1}\lambda_j^\ast
  c_{j+1}^\dagger c_j+\mathrm{h.c.}\right)P_{j+2}+
  |\lambda_j|^2P_{j-1}P_{j+1} \right]
  \label{eqn:hamiltonian}
\end{equation}
The first term is thus a hopping term that preserves the hard-core
exclusion. The second term can be interpreted as a potential, yielding
a contribution $|\lambda_j|^2$ whenever the sites $j-1$ and $j+1$ are
both empty. Since $Q^\pm$ change the fermion number by $\pm 1$, $H$
preserves the total number of fermions. In terms of operators, the
fermion number operator $F = \sum_{j=1}^N c_j^\dagger c_j$ obeys
$[Q^\pm,F]=\pm Q^\pm$ and $[Q^\pm, H]=0$.
The ensemble of algebraic relations between the four operators
$H,\,Q^\pm,\,F$ is the well-known $\mathcal N=2$ supersymmetry algebra
\cite{Witten}.

The supersymmetry has considerable consequences for the spectrum of
$H$. Some important ones are: \textit{(i)} all energies are positive or
zero $E\geq 0$, \textit{(ii)} all eigenstates with $E>0$ can be
organized into doublets of the supersymmetry algebra, \textit{(iii)}
an eigenstate has $E=0$ if and only if it is annihilated by both $Q^+$
and $Q^-$. Because of the latter property, finding the exact number of $E=0$
ground states is equivalent to computing the cohomology of $Q$, which
is the dimension of the space of states that are annihilated by $Q$
but cannot be written as $Q$ acting on something else \cite{Witten}.
This computation is simple to do in this one-dimensional case
\cite{fendley:02}.  The Hamiltonian (\ref{eqn:hamiltonian}) with
periodic boundary conditions has exactly two zero-energy ground states
when the number of sites is a multiple of three and all the
$\lambda_j$ obey $0<|\lambda_j|<\infty$. Moreover, the number of
fermions in both these ground states is $n\equiv N/3$. Henceforth we
restrict our analysis to $N=3n$ sites and study these ground states in
depth.

The case $\lambda_j=1$ for all sites $j$ was extensively studied in
\cite{fendley:02,fendley:03}. The model is solvable by means of the
coordinate Bethe ansatz. The Bethe equations are similar to those for
the XXZ model at $\Delta=-1/2$ with twisted boundary conditions
\cite{razumov:01,degier:02}, and indeed in given momentum sector a
mapping between the two models can be constructed
\cite{fendley:03,yang:04}.  The standard Bethe ansatz analysis of the
excitations above these ground states shows that the spectrum is
gapless, and so the model is critical in this case. The field theory
discribing the scaling limit is a simple free massless boson,
compactified at the supersymmetric radius and therefore the simplest
field theory with $\mathcal N=(2,2)$ supersymmetry (in the continuum
the supersymmetry is enhanced to two chiral and two anti-chiral
supersymmetries).

When all $\lambda_j=1$, the model possesses a translation symmetry.
Defining the translation operator $\mathcal T$ by ${\cal T}^{-1}c_j^\dagger {\cal T}=c^{\dagger}_{j+1}$, we have
$[H,{\cal T}]=0$. The cohomology computation shows that translation
invariance is spontaneously broken when the number of sites $N=3n$ is
a multiple of three: the eigenvalues of ${\cal T}$ on the two ground
states are $(-1)^n e^{\pm i\pi/3}$. Thus translation by three sites
${\cal T}^3$ on the lattice turns into translation symmetry in the
field theory describing the scaling limit. (In the field theory, the
eigenvalues of ${\cal T}$ in general are related to 
those of the ${\mathbb Z}_3$ subgroup of the chiral
fermion number \cite{Huijse}.)

Staggering the couplings typically destroys criticality in lattice
chains. Here this is the case as well. However, since the model
remains supersymmetric for any choice of the $\lambda_j$, it is
possible to preserve supersymmetry in the non-critical theory. 
We choose staggered coupling constants $\lambda_j$ with a period of three sites $\lambda_{j+3p} = \lambda_j$,
in order to keep the same translation symmetry in the scaling limit.
To preserve parity symmetry, i.e. invariance under reversal of the order of sites, we choose
\begin{equation}
\lambda_{3i}=\lambda_{3i+2}\equiv y, \qquad \lambda_{3i+1}=1\ .
\end{equation}
Defining  $P c_j P^{-1}
= c _{N-j+1}$, we have $[H, P]=0$ and $[{\cal T}^3, H]=0$.
For simplicity, we focus on the case of real coupling constants $y$. Then it
follows that in the staggered case the supercharges are linear
functions of the coupling constant $Q^\pm=Q_0^\pm+yQ_1^\pm$, and the
Hamiltonian becomes a quadratic polynomial $H = H_0+y H_1 + y^2 H_2$.

There is only one Lorentz-invariant supersymmetry-preserving
deformation of the free boson conformal field theory, the sine-Gordon
field theory at the supersymmetric point ($\beta^2=16\pi/3$ in
conventional normalization; for an overview see \cite{FI}). We thus
expect the scaling limit of this staggered lattice model to be
described by this massive field theory. This field theory is
integrable, and we expect that the quantum chain is as well. This
assertion is easy to check by finding the spectrum of $H$ as a
function of $y$ on the computer for finite sizes. One finds that the
characteristic polynomial of $H$ factorizes into polynomials in $y^2$
with integer coefficients, characteristic of an enhanced
symmetry. Moreover, there are many level crossings as $y$ is varied,
again characteristic of integrability.  Although we will not exploit
this integrability directly in our analysis, the special properties we
find for this model certainly stem from its presence.

\section{Properties of the ground states}
\label{sec:gswf}

Because we know the ground-state energy is exactly zero for any $N$,
it is natural to expect that the ground state itself will have special
properties. In a supersymmetric theory one expects even more, since
$Q^+$ and $Q^-$ individually annihilate the ground state. In this
section we describe first how this automatically requires some useful
properties of the ground states. We then show how these ground states
possess a number of much more surprising and elegant properties.

The supersymmetry means it is easy to find the quantum numbers of the
ground states for any value of the staggering. The cohomology
computation shows that the two ground states have eigenvalue of
$(-1)^{N+1}$ under
${\cal T}^3$, while one is parity even and the other parity odd. We
thus label the two ground states as
$|\Psi^{\rm e}(y)\rangle$ and $|\Psi^{\rm o}(y)\rangle$. When
the superscript is omitted, the equation applies to both ground states.

\subsection{Asymptotic limits}
\label{sec:asymptotic}

One nice feature of introducing the staggering is that the model can
easily be solved when $y\to 0$ and $y\to\infty$. This provides very
useful intuition into the general situation, and an example of the
power of supersymmetry.

In the limit $y\to 0$, the ground state must solve the equations
$Q_0^-|\Psi\rangle=Q_0^+|\Psi\rangle=0$. For $Q_0^-$ to annihilate
$|\Psi\rangle$ it must not contain any particles on the sites
$2,5,8,\dots$\ . For $Q_0^+$ to annihilates the ground state as well,
any such site must have a particle to its left or right; the hard-core
constraint then forbids creating a particle on these sites. There are
only two configurations with these two properties, each with $n=N/3$
fermions:
\begin{eqnarray*}
  |{\cal A}\rangle &=|\circ\circ\bullet\circ\circ\bullet\cdots\rangle = \left(\prod_{j=1}^n c_{3j}^\dagger\right) |0\rangle,\\
  P|{\cal A}\rangle &=|\bullet\circ\circ\bullet\circ\circ\cdots\rangle = \left(\prod_{j=1}^n c_{3j-2}^\dagger\right) |0\rangle.
\end{eqnarray*}
where $|0\rangle$ is the empty state, and to eliminate an annoying
sign we define $P|0\rangle=(-1)^{N+1}|0\rangle$. 
Parity eigenstates are given by
the linear combinations
\begin{equation}
|\Psi^{\rm e}(y=0)\rangle = |{\cal A}\rangle + P|{\cal A}\rangle,\qquad
|\Psi^{\rm o}(y=0)\rangle = |{\cal A}\rangle - P|{\cal A}\rangle\ .
\label{psi0}
\end{equation}

In the limit $y\to +\infty$, the ground state must be a solution of
$Q_1^-|\Psi\rangle=Q_1^+|\Psi\rangle=0$. We know that both $Q_1^-$ and
$Q_1^+$ annihilate and create fermions only on sites $1,3,4,6,\dots$
If all the sites $2,5,8,\dots$ are occupied, the hard-core constraint
means it is impossible to create or annihilate particles on any other
sites.  Thus the polarized state
\begin{equation*}
  |\Psi^{\rm e}(y\to\infty)\rangle \propto |{\cal P}\rangle=
 |\circ\bullet\circ\circ\bullet\circ\cdots\rangle = \left(\prod_{j=1}^n c_{3j-1}^\dagger\right) |0\rangle
\end{equation*}
is a ground state with
even parity in this limit.
In order to construct the parity-odd ground state, we notice
that any state with $n$ particles and all sites
$2,5,8,\dots$ empty contains exactly one particle on each pair of sites
$3j,3j+1$. It is therefore annihilated by
$Q_1^+$.  In order for the state to be
annihilated by $Q_1^-$ the particles on these two sites must occur as
``singlets'' $|\cdots \circ \bullet \cdots\rangle-|\cdots\bullet \circ
\cdots\rangle$, and thus the second ground state appears as tensor
product of such singlets
\begin{equation*}
  |\Psi^{\rm o}(y\to \infty) \propto
 |{\cal S}\rangle=
\prod_{j=1}^n\left(c_{3j+1}^\dagger-c_{3j}^\dagger\right)|0\rangle\  .
\end{equation*}

\subsection{Characterizing by the ground state by a polynomial}
\label{sec:polynom}

Since the ground states are annihilated by $H$ for all $y$, we can
imagine constructing the ground state by iteration. Namely, $H
= H_0+y H_1 + y^2 H_2$, where $H_0 = \{Q_0^+,Q_0^-\}$. We just saw
that $|\Psi(0)\rangle$ is annihilated by both $Q^+_0$ and $Q^-_0$, so
that $H_0 |\Psi(0)\rangle =0$.
Then we define 
\begin{equation}
|\Psi(y)\rangle \equiv |\Psi_0\rangle + y |\Psi_1\rangle + y^2
|\Psi_2\rangle+\dots
\label{Psiexp}
\end{equation}
where the $\Psi_i$ are independent of $y$. 
Since $H|\Psi(y)\rangle=0$, we therefore have 
\begin{eqnarray}
\nonumber
H_0|\Psi_1\rangle &=& - H_1|\Psi_0\rangle\\
H_0|\Psi_a\rangle &=& - H_1|\Psi_{a-1}\rangle - H_2
|\Psi_{a-2}\rangle\ .
\label{iterate}
\end{eqnarray}
for $a\ge 2$.
Thus if we could invert $H_0$, then by iterating we can determine any
$|\Psi_a\rangle$ from $|\Psi_0\rangle$.

Of course, $H_0$ is not invertible on the full Hilbert space, since
$H_0|\Psi_0\rangle=0$. However, it is easy to check that the states
given in (\ref{psi0}) are the unique
states annihilated by $H_0$ for a given parity. Thus let us write for
the $N=3n$-site system,
\begin{eqnarray*}
|\Psi(y)\rangle \equiv m(y) |\Psi_0\rangle +
|\widetilde{\Psi}\rangle,
\end{eqnarray*}
where $\langle \Psi_0 |\widetilde{\Psi}\rangle=0$. Then given
$m(y)$, this iteration procedure uniquely determines
$|\widetilde{\Psi}\rangle$. Moreover, since $H_0$ is diagonal, it is
easily inverted in the Hilbert space with $|\Psi_0\rangle$
excluded. 

The ground state can thus be characterized by this function $m(y)$.
We will conjecture a recursion relation for it in $n$ by noting the
connection to the XYZ chain and the Painlev\'e VI differential
equation. Because this analysis is technically somewhat involved, we
defer this calculation to section \ref{sec:xyz}.  However, several
important properties of $m(y)$ and the ground states follow solely
from the supersymmetry.

To illuminate these properties, it is useful to define the operator
\begin{equation}
S= \sum_{j=1}^n c^\dagger_{3j+1} c_{3j+1}\ 
\label{Sdef}
\end{equation}
that counts the number of fermions on the ``staggered'' sites $2,5,8,\dots$ Since
$Q_1^\pm$ does not change this number, $[S,Q_1^\pm]=0$,
but $[S,Q_0^\pm]=\pm Q_0^\pm$. Therefore $S$ commutes with $H_0$ and
$H_2$, but not $H_1$. The latter hops a particle on or off one of the
staggered sites, and so changes $S$ by $\pm 1$, so that
$\{(-1)^S,H_1\}=0$.  Thus $H(-y)= (-1)^S H(y) (-1)^S$. 

This observation means that $m(y)$ is a polynomial in $y^2$ when
appropriately normalized. Then in the expansion (\ref{Psiexp}) we have
$(-1)^S |\Psi_a\rangle = (-1)^a |\Psi_a\rangle$. States with even $a$
are therefore orthogonal to those with odd $a$, and so $m(y)$ can only
depend on $y^2$. For finite $n$, it is a finite polynomial; one needs
to iterate only a finite number of times to determine the full
wavefunction.  The expectation value of any operator that commutes
with $(-1)^S$ is therefore an even function of $y$, while any operator
that anticommutes is an odd function.

Let us give a few explicit examples. For any state $|\alpha\rangle$, define
its coefficient in the ground states by $\psi_\alpha\equiv \langle
\Psi^{\rm e} |\alpha\rangle $  and $\chi_\alpha \equiv \langle
\Psi^{\rm o} |\alpha\rangle $. For $N=6$ sites we find the components
$$
  \begin{array}{lp{3cm}l}
  \psi_{\circ\circ\circ\bullet\circ\bullet}=0 && \chi_{\circ\circ\circ\bullet\circ\bullet}=-2y^2\\
  \psi_{\bullet\circ\circ\bullet\circ\circ}=1 && \chi_{\bullet\circ\circ\bullet\circ\circ}=1+2y^2\\
  \psi_{\bullet\circ\circ\circ\bullet\circ}=-y && \chi_{\bullet\circ\circ\circ\bullet\circ}=-y\\
   \psi_{\circ\bullet\circ\circ\bullet\circ}=2y^2 &&  \chi_{\circ\bullet\circ\circ\bullet\circ}=0
  \end{array}
$$
For $N=9$ sites we have
$$
  \begin{array}{lp{0.7cm}l}
  \psi_{\circ\circ\circ\bullet\circ\bullet\circ\bullet\circ}=y^3 && \chi_{\circ\circ\circ\bullet\circ\bullet\circ\bullet\circ}=-y^3\\
  \psi_{\circ\circ\circ\bullet\circ\bullet\circ\circ\bullet}=-y^2&&
  \chi_{\circ\circ\circ\bullet\circ\bullet\circ\circ\bullet}=y^2(1+2y^2)\\
  \psi_{\bullet\circ\circ\bullet\circ\circ\bullet\circ\circ}=1+4y^2&&
  \chi_{\bullet\circ\circ\bullet\circ\circ\bullet\circ\circ}=1+2y^2+2y^4\\
  \psi_{\bullet\circ\circ\circ\bullet\circ\bullet\circ\circ}=-y(1+3y^2)&&
  \chi_{\bullet\circ\circ\circ\bullet\circ\bullet\circ\circ}=-y(1+y^2)\\
  \psi_{\bullet\circ\circ\circ\bullet\circ\circ\bullet\circ}=y^2(1+4y^2)&&
  \chi_{\bullet\circ\circ\circ\bullet\circ\circ\bullet\circ}=y^2\\
  \psi_{\bullet\circ\circ\circ\circ\bullet\circ\bullet\circ}=-2y^3&&
  \chi_{\bullet\circ\circ\circ\circ\bullet\circ\bullet\circ}=0\\
  \psi_{\circ\bullet\circ\circ\bullet\circ\circ\bullet\circ}=-2y^3(1+4y^2)&&
  \chi_{\circ\bullet\circ\circ\bullet\circ\circ\bullet\circ}=0
  \end{array}
$$ The remaining $\psi_\alpha$ and $\chi_\alpha$ are found by acting
with ${\cal T}^3$ and $P$ and arranging the states into appropriate
multiplets. Note that these coefficients are indeed even and odd in
$y$ depending on the value of $(-1)^S|\alpha\rangle$.  With our sign
convention $P|0\rangle=(-1)^{n+1}|0\rangle$, $P$ acting on all these
states with fermion number $n$ is found simply by inverting the
picture; no extra minus signs occur.  For example, $P|\bullet
\circ\circ\bullet\circ\circ\rangle =
|\circ\circ\bullet\circ\circ\bullet\rangle$.  Thus any state invariant
under inversion does not exist for odd parity.

\subsection{Special properties of the ground state}
\label{sec:special}

As discussed in the introduction, the ground states of these
supersymmetric models possess a number of remarkable properties.  We
have above detailed some properties that can be derived using the
supersymmetry. In this subsection we detail some special properties
that do not automatically follow from the supersymmetry. They
presumably are a consequence not only of the supersymmetry but the
underlying integrability of the chain. However, as with the analogous
results for the XYZ chain in
\cite{bazhanov:05,bazhanov:06,mangazeev:10,razumov:10}, these
properties are not derived but rather discovered by analyzing the
explicit ground states. 

We define the polynomials
characterizing the even- and odd-parity ground states with $n$
fermions and $N=3n$ sites by
\begin{eqnarray*}
m_n(y^2) &=& \langle \Psi^{\rm e} |\Psi^{\rm e}(0)\rangle/2,\\
m_{-n}(y^2) &=& \langle \Psi^{\rm o} |\Psi^{\rm o}(0)\rangle/2\ .
\end{eqnarray*}
These are normalized so that $m_n(0)=m_{-n}(0)=1$.  
Once these polynomials are known for a given $n$, the rest of the
ground state follows simply by the interaction procedure.

In fact, the following observation allows us to determine $m_{\pm n}$
iteratively on the computer using Maple, so that one does not need to
simultaneously iterate while looking for $m_{\pm n}$. Namely, with the
normalization that $m_{\pm n}(0)=1$, we find (up to $n=8$) the
surprising fact that the norm of the ground states is also given
simplify in terms of the $m_{\pm n}$ and $m_{\pm(n+1)}$:
\begin{eqnarray}
\nonumber
  \langle \Psi^{\rm e}(y) |\Psi^{\rm e}(y) \rangle
&=& 2m_{n}(y^2)\,m_{-(n+1)}(y^2)\ ,\\
\langle \Psi^{\rm o}(y) |\Psi^{\rm o}(y) \rangle 
&=& 2m_{-n}(y^2)\,m_{n+1}(y^2)\ .
\label{eqn:norms}
\end{eqnarray}
This means that we can determine the polynomials for the ground state
with $n+1$ fermions simply by computing the norm of the ground states
for $n$ fermions.
The first few polynomials are given by
\begin{eqnarray*}
&&  m_{1}(y^2)=m_2(y^2)=1, \qquad\quad
  m_{3}(y^2)=1+4y^2 ,\\
&& m_{4}(y^2)=1+5y^2+8y^4 , \qquad
  m_{5}(y^2)=1+11y^2+42y^4+80y^6+64y^8, \\
 &&m_{-1}(y^2)=1\ ,\qquad m_{-2}(y^2)=1+2y^2\ ,\qquad
m_{-3}(y^2)=1+2y^2+2y^4,\\&&
m_{-4}(y^2)=1+7y^2+12y^4+14y^6+8y^8,\\
&&  m_{-5}(y^2)=1+9y^2+30y^4+46y^6+54y^8+42y^{10}+16y^{12}\ .
\end{eqnarray*}

With the ground states in hand up to $n=8$ (so that we have $m_{\pm
  n}(y^2)$ up to $n=9$), we have made the following observations:
\begin{itemize}

\item 
All $\psi_\alpha$ and $\chi_\alpha$ are even or odd polynomials in $y$
with integer coefficients.  Note that the iteration procedure only
requires that the coefficients be rational; the fact that they are
integers is special.

\item All the integers in the polynomial for a given $\psi_\alpha$ or
  $\chi_\alpha$ have the same sign.
  These signs are opposite for any two configurations related by
  moving a single fermion by one site.

 \item The degree of $m_n(y^2)$ as a polynomial in $y$ for $n$ positive is 
 $\lfloor (n-1)^2/2\rfloor$, whereas the degree of $m_{-n}(y^2)$ is
 $\lfloor n^2/2\rfloor$, where $\lfloor a \rfloor$ is the largest
 integer less than or equal to $a$.

   \item The polynomials $m_n$ appear elsewhere in the ground
     state. Namely, the projection of the parity-even ground state onto the
   fully-polarized state (its ground state at $y\to \infty$) is
\begin{equation}
\langle {\cal P}|\Psi^{\rm e}(y)\rangle = 2y^n m_n(y^2)
 \label{eqn:projpol}
\end{equation}
Moreover, the projection on states that are fully polarized except for
one particle is $-y^{n-1} m_n(y^2)$. This property can directly be
confirmed by perturbation theory around the point $y=\infty$.
\item
The projection of the parity-odd ground state onto the
singlet state is
\begin{equation*}
     \langle {\cal S}|\Psi^{\rm o}(y)\rangle = 2 m_{n+1}(y^2).
\end{equation*}

\item When $\alpha$ is a ``fully squeezed'' state like
$\bullet\circ\bullet\circ\bullet\circ\circ\circ\circ$, 
$\chi_\alpha$ and $\psi_\alpha$ are monomials.

\item The polynomials $m_n(y^2)$ factorize into two polynomials in
  $y^2$ with integer coefficients when $n$ is odd, while $m_{-n}(y^2)$
  factorizes when $n$ is even. For example, $m_5(y^2) =
  (1+2y^2)(1+9y^2+24y^4+32y^6)$ and $m_{-4}(y^2)= (1+y^2) (1+6 y^2+6
  y^4+8 y^6)$. Below we will give a change of variables that will
  generalize the factorization to all $n$.

\item Evaluating the polynomials at the critical point $y=1$ gives
  integers with interesting combinatorial properties \cite{Beccaria},
  very similar to what occurs in the XXZ chain at $\Delta=-1/2$ (see e.g.\
  \cite{razumov:01}). We detail some of these properties in the
  appendix.

\end{itemize}

\section{Duality and scaling of the order parameters}
\label{sec:dualityscaling}

In our earlier paper \cite{fendley:10} we described how to find simple
but exact expressions for several expectation values, and noted that
there seemed to be a duality between large and small $y$, exact even
on the lattice. In this section we review and expand upon these
results. In particular, we find a useful change of variable that
will allow us in the next section to find an explicit recursion
relation for the $m_{\pm n}(y^2)$.

\subsection{Order parameters}

There are three distinct one-point expectation values in this
supersymmetric chain. All three play the role of order parameters,
distinguishing the weak-coupling ($y<1$) phase from the
strong-coupling ($y>1$) phase. The staggered densities in the even-
and odd-parity ground states are the expectation values of the
operator $S$ from (\ref{Sdef}):
\begin{eqnarray*}
\rho^{\rm e}_n(y^2) = \frac{\langle \Psi^{\rm e}| S | \Psi^{\rm
  e}\rangle}{n \langle \Psi^{\rm e}| \Psi^{\rm   e}\rangle}
,\qquad
\rho^{\rm o}_n(y^2) = 
 \frac{\langle \Psi^{\rm o}| S | \Psi^{\rm
  o}\rangle}{n \langle \Psi^{\rm o}| \Psi^{\rm   o}\rangle}\ .
\end{eqnarray*}
Using the derivation in section \ref{sec:polynom}, these must be
functions of $y^2$. Our observation in section \ref{sec:special} implies that
these are ratios of polynomials in $y^2$ with integer coefficients.

{}From the asymptotic ground states found in section
\ref{sec:asymptotic} it is easy to see that 
\begin{eqnarray*}
\rho^{\rm e}_n(0)=\rho^{\rm o}_n(0)=\rho^{\rm o}_n(y^2=\infty)=0,
\qquad 
\rho^{\rm e}_n(\infty)=1\ .
\end{eqnarray*}
Translation invariance at the
critical point $y=1$ constrains the expectation values, because here the
combinations $| \Psi^{\rm
  e}\rangle \pm | \Psi^{\rm
  o}\rangle$ are eigenstates of the translation operator ${\cal
  T}$. The eigenvalue of $S$ in these eigenstates must be $n/3$, so
this requires
\begin{eqnarray*}
\rho^{\rm e}_n(1)+\rho^{\rm o}_n(1)=2/3\ .
\end{eqnarray*}
However, it does not follow that the difference of the two is zero at the
critical point: $S$ and ${\cal T}$ do not commute, so $S$ has
off-diagonal matrix elements between translation eigenstates. 
In fact, it not zero, as we will detail in section \ref{sec:scaling}
below.

In \cite{fendley:10} we described how various expectation values were
{\em scale free}. This means that the coefficients in the perturbative
expansion around completely ordered points $y=0,\infty$ are
independent of $N$ up to order $y^{\pm 2n}$. Let us illustrate this
here with the expansion of $\rho^{\rm e}(y^2) + \rho^{\rm o}(y^2)$ around
$y^2=\infty$ using the exact ground states:
\begin{eqnarray*} 
\rho_2^{\rm e}(y^2) + \rho_2^{\rm o}(y^2)&=&
1-{\frac {3}{8}}{y}^{-2}-{\frac {3}{32}}{y}^{-4}+{\frac 
{39}{128}}{y}^{-6}-{\frac {143}{512}}{y}^{-8}+{\frac {287}{2048}}{y}^{10
}+O \left( {y}^{-12} \right) \\
\rho_3^{\rm e}(y^2) + \rho_3^{\rm o}(y^2)&=&
1-{\frac {3}{8}}{y}^{-2}+{\frac {3}{64}}{y}^{-4}-{\frac 
{103}{512}}{y}^{-6}+{\frac {2203}{4096}}{y}^{-8}-{\frac {20623}{32768}}{
y}^{-10}+O \left( {y}^{-12} \right) \\
\rho_4^{\rm e}(y^2) + \rho_4^{\rm o}(y^2)&=&
1-{\frac {3}{8}}{y}^{-2}+{\frac {3}{64}}{y}^{-4}-{
\frac {3}{512}}{y}^{-6}-{\frac {611}{2048}}{y}^{-8}+{\frac {16291}{16384
}}{y}^{-10}+O \left( {y}^{-12} \right) \\
\rho_5^{\rm e}(y^2) + \rho_5^{\rm o}(y^2)&=&
1-{\frac {3}{8}}{y}^{-2}+{\frac {3}{64}}{y}^{-4}-{
\frac {3}{512}}{y}^{-6}+{\frac {3}{4096}}{y}^{-8}-{\frac {15879}{32768}}
{y}^{-10}+O \left( {y}^{-12} \right) \\
\rho_6^{\rm e}(y^2) + \rho_6^{\rm o}(y^2)&=&
1-{\frac {3}{8}}{y}^{-2}+{\frac {3}{64}}{y}^{-4}-{
\frac {3}{512}}{y}^{-6}+{\frac {3}{4096}}{y}^{8}-{\frac {3}{32768}}{y}^
{-10}+O \left( {y}^{-12} \right) 
\end{eqnarray*}
It is thus natural to conjecture that this pattern persists for all
$N$, and so we can sum the series to give
\begin{equation}
\rho_\infty^{\rm e}(y^2) + \rho_\infty^{\rm o}(y^2) = 
\frac{8y^2-2}{1+8y^2}\ ,\qquad y\ge 1.
\label{asymhigh}
\end{equation}
The fact that this expression gives the correct value of $2/3$ at the
critical point $y=1$ is a strong indication that this formula is both
exact and valid for all $y\ge 1$. The expansion around $y=0$ has the same scale free property, and summing this series gives
\begin{equation}
\rho_\infty^{\rm e}(y^2) + \rho_\infty^{\rm o}(y^2) = 
\frac{4y^2}{1+2y^2+\sqrt{1+8y^2}}\ ,\qquad{y\le 1}
\label{asymlow}
\end{equation}
The two expressions are indeed continuous at $y=1$, but
the second derivatives are different, as one would expect at a critical point.

Another order parameter involves the fermion densities on the
sites $1,3,4,6,\dots$ Because overall fermion number is conserved, it is easy to see that
\begin{eqnarray*}
F = S + {\cal T} S {\cal T}^{-1} + {\cal T}^{-1} S {\cal T}\ 
\end{eqnarray*}
is a constant.
Therefore to get something new, we should look at the difference of
the latter two operators. The difference is odd under parity, so it
maps one ground state to the other. We define
\begin{equation}
\tau_n(y^2) = 
\frac{\langle \Psi^{\rm e}| 
\left({\cal T} S {\cal T}^{-1}-{\cal T}^{-1} S {\cal T}\right)
 | \Psi^{\rm  o}\rangle}{n 
\sqrt{\langle \Psi^{\rm e}| \Psi^{\rm   e}\rangle
\langle \Psi^{\rm o}| \Psi^{\rm o}\rangle}}\ .
\label{eqn:poddorder}
\end{equation}
Rewriting this in terms of translation eigenstates at the critical point
and exploiting the fact that the translation eigenvalues there are
$(-1)^n e^{\pm i\pi /3}$ gives
\begin{eqnarray*}
\tau_n(1) = \frac{\sqrt{3}}{2}
\left(\rho^{\rm e}_n(1) - \rho^{\rm o}_n(1)\right)\ .
\end{eqnarray*}
The order parameter $\tau_n$ as well as the individual $\rho^{\mathrm e}_n,\,\rho^{\mathrm o}_n$ are scale free as well, and summing their
expansions as above gives conjectures analogous to \eqref{asymhigh} and \eqref{asymlow}. We will give their
expressions below, after finding a useful reparametrization of $y$.

\subsection{Duality}
\label{sec:duality}

Several considerations suggest that there is a duality between strong
($y>1$) and weak ($y<1$) coupling. Because the sine-Gordon field theory
is the same for either sign in front of the off-critical
perturbation, the scaling limit of this lattice model should be the
same for $y$ above and below the critical point $y=1$. Moreover, as we
will detail in section \ref{sec:xyz}, the XYZ chain in the same
universality class possesses an explicit duality symmetry. We will
show in this subsection that there indeed is exact duality relation
between certain combinations of the order parameter expectation values
in our fermion chain. This duality relation is exact even for a finite
number of sites.

To exhibit this duality, it is very convenient to parametrize the
staggering by a different variable. Examining the exact expression
(\ref{asymhigh}) suggests that we find a variable that removes the awkward
square roots. An obvious choice is $\zeta=\sqrt{1+8y^2}$, and indeed
in the next section we will utilize this variable, as it makes the
connection to the XYZ chain transparent. To understand the duality, $\zeta$
is less convenient, because the critical self-dual point occurs at
$\zeta=3$. Thus in this section, we utilize the variable $v$ defined
by
\begin{equation}
y^2=\frac{1-v}{(1+v)^2}\ .
\label{yv}
\end{equation}
The critical point occurs at $v=0$, with the $y>1$ regime
corresponding to $-1<v<0$ and the $0\le y<1$ regime corresponding to
$1\ge v>0$. We will show that the duality
transformation is simply $v\to -v$.

To find which quantities are invariant under the duality
transformation, let us first
rewrite the asymptotic expressions for the order parameters in terms
of $v$ using (\ref{yv}). By slight abuse of notation, we will keep the same symbols, 
i.e. write $\rho_n^{\mathrm e}=\rho_n^{\mathrm e}(v)$. The new parametrization allows us to combine
(\ref{asymhigh}) and (\ref{asymlow}) into a single expression valid in
both regimes:
\begin{equation}
\rho_\infty^{\rm e}(v) + \rho_\infty^{\rm o}(v) = 
\frac{2}{3} - \frac{8}{3}\frac{v(v+3)}{(3+|v|)^2}\ .
\label{rhoinf}
\end{equation}
Obviously, this is not invariant under $v\to -v$. However, it does
suggest that we study the quantity
\begin{equation*}
\beta_n(v)\equiv \frac{1}{v(v+3)}
\left(\rho_n^{\rm e}(v)+ 
\rho_n^{\rm o}(v)-\frac{2}{3}\right)\ .
\end{equation*}
By analyzing the exact expectation values for $n$ up to 8, we
find that $\beta_n(v)$ is invariant under duality for finite values of
$n$ as well:
\begin{eqnarray*}
\nonumber
\beta_n(v)=\beta_n(-v)\ .
\end{eqnarray*}
Thus it is natural to conjecture that $\beta_n$ is a function of $v^2$
for all $n$. Of course, if one reverts to the original parameter $y$,
it remains a function solely of $y^2$ as well.

The other expectation values discussed above also have nice
expressions in terms of $v$. By summing the scale-free series, we have
\begin{equation}
\rho_\infty^{\rm e}(v) - \rho_\infty^{\rm o}(v) = 
\frac{8\sqrt{v(v+3)(v-1)}}{(v-3)^2},\qquad 0\ge v\ge -1
\label{rhodiffinf}
\end{equation}
while $\rho_\infty^{\rm e}(v) - \rho_\infty^{\rm o}(v)=0$ for $1>v>0$.
Combining this with (\ref{rhoinf}) gives for the product 
\begin{eqnarray*}
\rho_\infty^{\rm e}(v) \rho_\infty^{\rm o}(v) = \left(\frac{1-|v|}{3+|v|}\right)^2
\end{eqnarray*}
for all $v$. This suggests we examine 
\begin{equation*}
\gamma_n(v)\equiv \rho_n^{\rm e}(v)
\rho_n^{\rm o}(v)\ .
\end{equation*}
We have checked up to $n=8$ that 
this quantity is indeed self-dual: $\gamma_n(v)=\gamma_n(-v)$.

For the off-diagonal order parameter defined in \eqref{eqn:poddorder}, we sum the series to get 
\begin{equation}
\tau_\infty(v) = 
\frac{4\sqrt{v}}{v+3},\qquad 0\le v\le 1
\label{tauinf}
\end{equation}
while $\tau_\infty=0$ for $0\ge v \ge  -1$.
This has a square-root singularity at the critical point 
similar to (\ref{rhodiffinf}), and so suggests that
we examine the product
\begin{equation*}
\delta_n(v)\equiv 
\left(\rho_n^{\rm e}(v) - \rho_n^{\rm o}(v)\right)\tau_n(v) \ .
\end{equation*}
We have checked up to $n=8$ that this quantity
is indeed self-dual:
$\delta_n(v)=\delta_n(-v)$.

\subsection{Critical exponents and scaling functions}
\label{sec:scaling}

We can extract various critical exponents from the exact expressions
we have found. These match what is expected from conformal field
theory, and so provide additional evidence that our conjectures for
the $N\to\infty$ limit are correct. Moreover, they allow us to define
appropriate off-critical {\em scaling functions}. A scaling function
has a well-defined limit as $N\to\infty$. To find one in an
interacting system, the ``bare'' parameters (those in the Hamiltonian)
must be rescaled. Here we will show that the appropriate rescaling is
to plot quantities in terms of $n^{2/3}v$.

Let us start with the behavior precisely at the critical point. 
As we showed above, $\rho^{\rm e}_n(v=0)+\rho^{\rm o}_n(v=0)=2/3$
exactly for any $n$.  
Using the exact ground states, we find that for
$n=2,3,4,5,6,7,8$, $\rho^{\rm e}_n(v=0)-\rho^{\rm o}_n(v=0)$
is
$$
{\frac {8}{15}},{\frac {7}{15}},{\frac {14}{33}},{\frac {13}{33}},{
\frac {208}{561}},{\frac {988}{2805}},{\frac {1976}{5865}}
$$
A little trial and error shows that this sequence is given by
\begin{equation*}
\rho^{\rm e}_n(v=0)-\rho^{\rm o}_n(v=0)\ =\  
\frac{4}{3}\prod_{k=1}^n \frac{3k-2}{3k-1}\ =\ 
\frac{4}{3}\frac{\Gamma(2/3)\Gamma(1/3+n)}{\Gamma(1/3)\Gamma(2/3+n)}\ .
\end{equation*}
If we assume this applies for all $n$ and then use Stirling's formula,
we find that for large $n$
\begin{eqnarray*}
\rho^{\rm e}_n(v=0)-\rho^{\rm o}_n(v=0)\ =\
n^{-1/3}\frac{4\Gamma(2/3)}{3\Gamma(1/3)}\left(1-\frac{1}{3n}+\dots\right).
\end{eqnarray*}
The fact that this vanishes as $n^{-1/3}$ indicates that in conformal
field theory this is the expectation value of an operator with
dimension $1/3$. There indeed is an operator of this dimension in the
superconformal field theory identified in
\cite{fendley:02,fendley:03,Huijse} as describing the scaling limit at
the critical point.

Going off the critical point, we see that both $\rho^{\rm
e}_\infty(v)-\rho^{\rm o}_\infty(v)$ and $\tau_\infty(v)$ have square-root
singularities in $v$ as the critical point $v=0$ is approached. Since at the
critical point, both quantities scale with system size as $n^{-1/3}$,
the standard scaling argument indicates that the operator perturbing
the theory away from the critical point has dimension $4/3$. This
argument is as follows. Perturbation theory around the critical point
gives
\begin{eqnarray*}
\tau_n (v) \approx A(n)\left(1+a_1(n) v + a_2(n) v^2 + \dots\right)\ .
\end{eqnarray*}
For $n$ large we have $A(n)\propto 1/n^{1/3}$, and for $v$ small we
have $\lim_{n\to\infty}\tau_n(v)\propto \sqrt{v}$. The only way this
is consistent for $n$ large is for the coefficients
$a_k(n)$ to diverge as $n^{2k/3}$. To ensure an appropriate scaling
limit, we define a rescaled variable $w\equiv n^{2/3}v$ that we keep
finite as $n\to\infty$. This variable $w$ has scaling dimension $2/3$, so if
we consider an effective Hamiltonian near the critical point
$H=H_{crit}+w\int \diff x\, \Phi(x)$, the operator $\Phi(x)$ must have
dimension $4/3$. This indeed is the dimension of the only relevant
supersymmetry-preserving deformation of this superconformal field
theory  \cite{FI}.

\begin{figure}[t] 
\begin{center} 
\includegraphics[width= .4\textwidth]{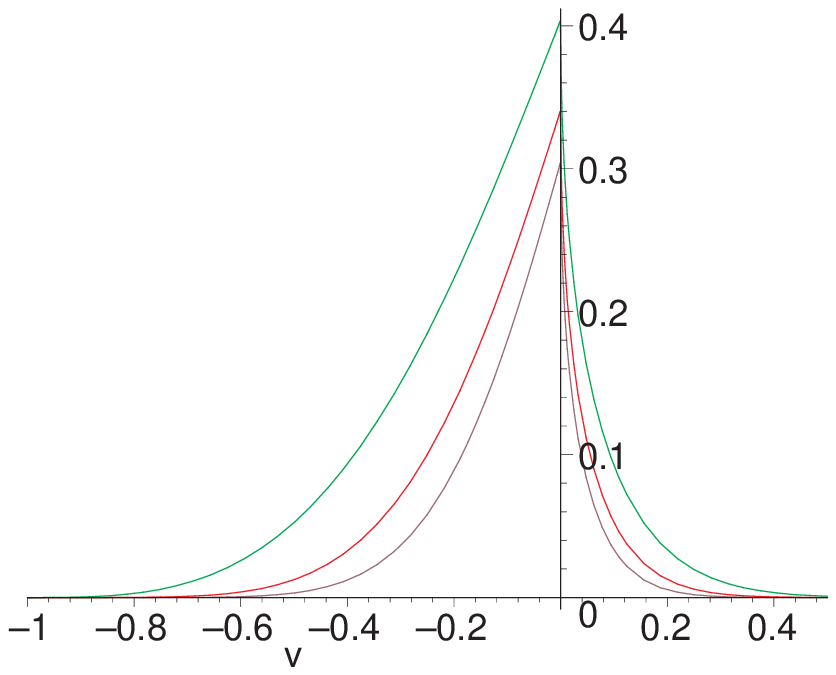} 
\qquad
\includegraphics[width= .4\textwidth]{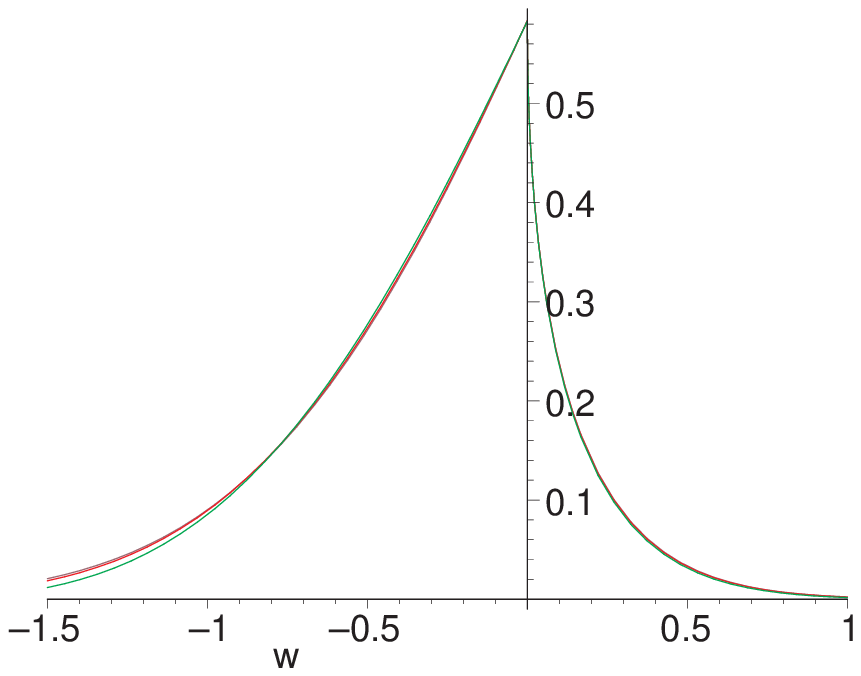} 
\caption{On the left, $\tau_n(v)-\tau_\infty(v)$ as a function of $v$ for
  $n=3,5,7$. On the right, the rescaled version $g(w)$ defined in
  (\ref{gdef}).}
\label{fig:tau} 
\end{center} 
\end{figure}

Knowing these exponents, we can now find scaling functions that remain
finite and converge quickly in the critical region as
$n\to\infty$. Instead of depending on $n$ and $v$ independently, they
are functions of the single variable $w=n^{2/3}v$.  For example, we
define
\begin{equation}
g(w) = \lim_{n\to\infty} n^{1/3}\left(\tau_n(n^{-2/3}w)-\tau_\infty(n^{-2/3}w)\right)\ .
\label{gdef}
\end{equation}
To illustrate this, we plot $(\tau_n-\tau_\infty)$ as a function of
$v$ on the left of figure \ref{fig:tau} for sizes $9,15,21$.  Although
these go to zero in the trivially solvable limits $v=-1$ and $v=1$, we
see the effects of finite size strongly at the critical point
$v=0$. When we plot instead the rescaled version $g(w)$ on the right
side of figure \ref{fig:tau}, we see an almost-instant collapse to a single
curve. Thus we obtain a numerically-accurate scaling function from a
very small system size. This strong collapse almost certainly is a
consequence of the supersymmetry.

\section{Exact properties of the polynomials characterizing the ground state}
\label{sec:xyz}

In this section we explain how the simple change of variables from
$y$ to $\zeta=\sqrt{1+8y^2}$ makes obvious a direct relation to
polynomials that were studied by Bazhanov and Mangazeev
\cite{bazhanov:05,bazhanov:06,mangazeev:10}, and Razumov and Stroganov
\cite{razumov:10}, in the framework of a particular XYZ chain. They
turn out to be related to classical integrable equations and the
Painlev\'e VI transcendent. This relation proves to be very fruitful,
because several properties of the our chain can be deduced from
already-known features of this XYZ chain. Here we discuss how the
polynomials $m_{\pm n}(y^2)$ characterizing the ground state satisfy a recursion
relation in $n$. Like the order parameters discussed above,
they are scale free and obey a duality. The scale-free property in
particular allows us to find an explicit expression for their
$n\to\infty$ behavior.

\subsection{XYZ chain and the Painlev\'e VI tau-function hierarchy}
Let us first recall some facts about the particular XYZ chain which
possesses some striking properties \cite{baxtersolving,
bazhanov:05,bazhanov:06,mangazeev:10,razumov:10,fendley:10}.
The XYZ Hamiltonian is
\begin{equation}
  H_{\mathrm{XYZ}} = -\frac{1}{2}\sum_{k=1}^L
  J_x\sigma_k^x\sigma_{k+1}^x+J_y\sigma_k^y\sigma_{k+1}^y
 +J_z\sigma_k^z\sigma_{k+1}^z\ ,
    \label{eqn:xyzhamiltonian}
\end{equation}
where the Pauli matrices act on a the two dimensional Hilbert space at
each site. As above, we take periodic boundary conditions.
The special properties occur along the line of couplings
$J_xJ_y+J_xJ_z+J_yJ_z=0$.  A convenient parametrization
of the coupling constants along this line is given by
\begin{equation}
J_x =1+\zeta,\,J_y=1-\zeta,\,J_z = (\zeta^2-1)/2\ .
  \label{eqn:couplings}
\end{equation}
This model is an off-critical deformation of the XXZ model at
$\Delta=J_z/J_x=-1/2$ (the $\zeta=0$ case). In the conventions of our earlier
paper \cite{fendley:10}, $\zeta=3/s$.

An obvious symmetry of the XYZ spectrum comes by permuting the Pauli
matrices. Along the line (\ref{eqn:couplings}), this results in 
two useful transformations. A rotation
of all spins by $\pi/2$ around the $z$-axis exchangs the
$\sigma^x\sigma^x$- and $\sigma^y\sigma^y$-terms in
$H_{\mathrm{XYZ}}$. This is equivalent to the
transformation $\zeta\to -\zeta$. Conversely, rotating all spins by $\pi/2$ around
the $y$-axis exchanges the $\sigma^x\sigma^x$- and
$\sigma^z\sigma^z$-terms. Equivalently this can be done by the homographic
transformation
\begin{equation}
  \zeta \to \frac{\zeta+3}{\zeta-1},
  \label{eqn:homographic}
\end{equation}
and rescaling energy by $(\zeta-1)^2/4$. Both these transformations are
symmetries of the spectrum of $H_{XYZ}$, up to an overall rescaling.

When the number of sites $L$ is odd, all the eigenvalues of
$H_{\mathrm{XYZ}}$ are doubly degenerate under spin reversal. As
$H_{\mathrm{XYZ}}$ commutes with both the operators $I =
\prod_{k=1}^L\sigma^z_k$ and the spin-reversal operator $R =
\prod_{k=1}^L\sigma^x_k$ one may decompose the corresponding subspaces
into eigenfunctions with
$I|\Phi_\pm(\zeta)\rangle=\pm|\Phi_\pm(\zeta)\rangle$. These are
related by spin-reversal
$R|\Phi_\pm(\zeta)\rangle=|\Phi_\mp(\zeta)\rangle$, and thus one may
concentrate on one of them, say $|\Phi_-(\zeta)\rangle$.

Let us now turn to the ground states of the model. For an odd number of
sites $L=2n+1$ the ground states $|\Phi_\pm(\zeta)\rangle$ have energy
$E_0=-L(\zeta^2+3)/4$ exactly
\cite{baxtersolving,stroganov:01_2}. Like for our fermion chain,
consider a decomposition of the wave function into spin-configuration
states
$|\Phi_-(\zeta)\rangle=\sum_{\alpha}\phi_\alpha(\zeta)|\alpha\rangle$. Given
that the Hamiltonian itself and the ground state energy are quadratic
polynomials in $\zeta$, one can choose the normalization in such a way
that all components are polynomials in $\zeta$. An overall constant
normalization can be fixed by demanding that the zero-order term for
configurations $\tilde \alpha =\,
\uparrow\uparrow\cdots\uparrow\downarrow\downarrow\cdots\downarrow$
containing a block of $n$ (resp. $n+1$) spins up for $n$ even
(resp. odd) equals one. Then it turns out that all polynomials
$\phi_\alpha(\zeta)$ have integer coefficients and definite
parity. Some of them enjoy special properties: the component in front
of the fully-polarized state $\,\downarrow\downarrow\cdots\downarrow$
with all spins down is given by
\begin{equation}
  \phi_{\downarrow\downarrow\cdots\downarrow}(\zeta)=\zeta^{n(n+1)/2}s_n(\zeta^{-2})
  \label{eqn:polcompxyz}
\end{equation}
where the $s_n(z)$ are polynomials in $z$ of degree $\lfloor n^2/4\rfloor$ that solve the non-linear recursion relation
\begin{eqnarray}
  2z(z-1)(9z-1)^2(\ln s_n(z))'' + 2(3z-1)^2(9z-1)(\ln s_n(z))'\nonumber \\
+8(2n+1)^2 \frac{s_{n+1}(z)s_{n-1}(z)}{s_n(z)^2}
  -(4(3n+1)(3n+2)+(9z-1)n(5n+3))=0
  \label{eqn:hirota}
\end{eqnarray}
with initial conditions $s_0(z)=s_1(z)\equiv 1$. In fact, these are a
special case for the Hirota equations for the $\tau$-function
hierarchy corresponding to Picard solutions of the Painlev\'e VI
differential equation \cite{okamoto:87,mangazeev:10_2}. Moreover,
these polynomials also appear in the computation of the $Q$-function
for the ground state of the corresponding eight-vertex model.

There is a number of remarkable properties of the XYZ chain involving
the polynomials $s_n(z)$; for an extensive list see
\cite{mangazeev:10,razumov:10}.  As an example let us consider the
norm of the wavefunction. Finite-size computations show that it is
given in terms of the polynomials $s_n(z)$ by
\begin{equation}
  ||\Phi_-(\zeta)||^2 = (4/3)^n \zeta^{n(n+1)}s_n(\zeta^{-2})s_{-n-1}(\zeta^{-2}).
  \label{eqn:normxyz}
\end{equation}
Using the invariance of the ground state sector under
\eqref{eqn:homographic} it is not difficult to show that
\eqref{eqn:normxyz} is invariant under the homographic transformation
modulo rescaling by a $\zeta$-dependent function. After a little
algebra, one finds the invariant
\begin{equation}
 s_n(\zeta^{-2})s_{-n-1}(\zeta^{-2})=\left(\frac{3+\zeta}{2\zeta}\right)^{n(n+1)}s_n\left(\left(\frac{\zeta-1}{\zeta+3}\right)^2\right)s_{-n-1}\left(\left(\frac{\zeta-1}{\zeta+3}\right)^2\right).
 \label{eqn:invariants}
\end{equation}

\subsection{The polynomials $m_{\pm n}(y^2)$ and Painlev\'e VI}

We noted in section \ref{sec:special} that $m_n(y^2)$ factorizes into a
product of polynomials in $y^2$ with integer coefficients when $n$ is
odd, and $m_{-n}(y^2)$ factorizes when $n$ is even. However, as we
detail in the appendix, they factorize for all $n$ at the critical
point $y=1$. Thus one might hope that they factorize for all $y$ with
an appropriate change of variable, and indeed changing to $\zeta$ or
$v$ allows this factorization.

This change of variables allows us to find a much more
dramatic result: the $m_{\pm n}(y^2)$ are simply related to the
analogous polynomials $s_n(z)$ in the XYZ chain! 
By examining their explicit form up to $n=8$ in terms the
new variable $\zeta$,  it is readily apparent
that they are related to the $s_n(z)$ by the following
equations:
\begin{eqnarray}
 m_n\left(y^2=(\zeta^2-1)/8\right) &=&\
 \frac{\zeta^{\lfloor(n-1)^2/2\rfloor}}{2^{(n-1)(n-2)/2}}
 \frac{s_{n-1}(\zeta^{-2})}{s_{n-1}(0)},  \label{eqn:mpbp1}
\\ 
m_{-n}\left(y^2=(\zeta^2-1)/8\right) &=&\ 
 \frac{\zeta^{\lfloor
 n^2/2\rfloor}}{2^{(n-1)(n+2)/2}}\frac{s_{-n}(\zeta^{-2})}{s_{-n}(0)}.
 \label{eqn:mpbp2}
\end{eqnarray}
with $s_n(0)=1$ for $n\geq 0$, and $s_n(0) = (3/4)^{-n-1}$ for
$n<0$. 

The most important consequence of this identification is that the
polynomials $m_{\pm n}(y^2)$ can be obtained from a recursion
relation. Moreover, the ground states in the fermion and XYZ chains
have analogous special properties: for example, \eqref{eqn:polcompxyz}
is analogous to \eqref{eqn:projpol}. 

We can use the knowledge about properties of the $s_n(z)$ in order to
obtain ground state properties of our fermion chain. When the
transformation (\ref{eqn:homographic}) is rewritten in
terms of $v=(3-\zeta)/(\zeta+1)$, this is precisely the $v\to -v$
duality discussed in section (\ref{sec:duality}).
The
square norm of the ground state wavefunctions of the fermion chain, given in
\eqref{eqn:norms}, are quite similar, although not identical to
\eqref{eqn:normxyz}. However considering the combination of
\eqref{eqn:norms}, \eqref{eqn:mpbp1} and \eqref{eqn:mpbp2}, it is a straightforward
consequence of \eqref{eqn:invariants} that the product of the
ground state square norms $N_n (v)= ||\Psi^{\rm
e}(y)||^2||\Psi^{\rm o}(y)||^2$ has the following invariance property
\begin{equation}
  N_n\left(-v\right) = \left(\frac{v+1}{2}\right)^{n(n+1)}N_n(v).
\end{equation}
Given this invariance property it is not surprising that all quantities studied in section \ref{sec:duality} with nice duality properties involve both parity sectors.

\subsection{Scale-free properties}

Knowing that the $m_{\pm m} (y^2)$ are related to the Painlev\'e VI
equation teaches us a great deal about the ground states of our
fermion model. Here we show how the lessons we have learned in the
analysis of the fermion model give new insight into the polynomials
$s_n(x)$. In particular, we show how they obey the scale-free property
discussed in \cite{fendley:10} and above in section \ref{sec:duality}.
Combining this with the Hirota equations \eqref{eqn:hirota} 
allows us to derive the
scale-free coefficients explicitly. We can then sum this series by approximating
the Hirota equation, and so find an asymptotic expression for the
$s_{n}$ themselves.

To this end, let us introduce another reparametrization $x =
(1-\zeta)/(3+\zeta)$. Then define a rescaled version of $s_n$ via
\begin{equation}
  S_n(x)= s_n\left(1/\zeta^2\right)
(1+x)^{\frac{n(n+1)}{2}}(1-x)^{\frac{n(n+2)}{2}}(1+3x)^{\frac{n^2}{2}}/s_n(1)
  \ .
\end{equation}
where $s_n(1)=2^{n(n-1)/2}$ for $n\geq 0$, and
$s_n(1)=3^{-n-1}2^{(n+2)(n+1)/2}$ for $n<0$.  The Taylor series
expansion of $S_n(x)$ around the trivially solvable point $x=0$ is scale free:
for $n> 0$ the first $n$ terms
in this expansion are independent of $n$, while for $n<0$ the first $-n-1$
terms in this expansion are independent of $n$. The scale-free part of the expansion is
$$
  S(x) = 1+x+2 x^3-x^4+9 x^5-6 x^6+54 x^7-36 x^8+\dots
$$ 

We can determine all the coefficients and a summed form
directly from \eqref{eqn:hirota}:
the $S_n(x)$'s are solution to the recurrence equation
\begin{eqnarray*}
  p(x)(\ln S_n(x))'' &+ q(x)(\ln S_n(x))' 
+2(2n+1)^2\left(\frac{S_{n+1}(x)S_{n-1}(x)}{S_{n}(x)^2}-1\right)= r(x),
\end{eqnarray*}
with the coefficients
\begin{eqnarray*}
  p(x) &= \frac{2x(x-1)(x+1)^2(3x-1)^2}{1+3x},\\
  q(x) &= \frac{(x+1)(3x-1)(1+8x-22x^2+45x^4)}{(1+3x)^2},\\
  r(x) &= -3x^3-x^2+3x-1.
\end{eqnarray*}
Let us now make the assumption that the limit $S_\infty(x)=\lim_{n\to \infty}
S_n(x)$ exists and that we can neglect the term $\propto (2n+1)^2$
(i.e. the ratio $S_{n+1}(x)S_{n-1}(x)/S_{n}(x)^2-1$ decays faster than
$1/n^2$, at least for small $x$). Thus we must solve the differential
equation
\begin{equation}
  p(x)(\ln S_\infty(x))'' + q(x)(\ln S_\infty(x))'=r(x)
\end{equation}
with initial conditions $S_\infty(x=0)=1,\, S_\infty'(x=0) = 1$. The
unique solution is given by the simple function
\begin{equation}
  S_\infty(x) =
  \frac{(1+3x)^{5/24}}{(1-x)^{3/8}(1+x)^{1/8}(1-3x)^{1/24}}
\label{Sinf}
\end{equation}
which indeed generates all the scale-free coefficients.

\begin{figure}
  \centering
  \begin{tabular}{cp{0.075\textwidth}c}
\includegraphics[width=0.4\textwidth]{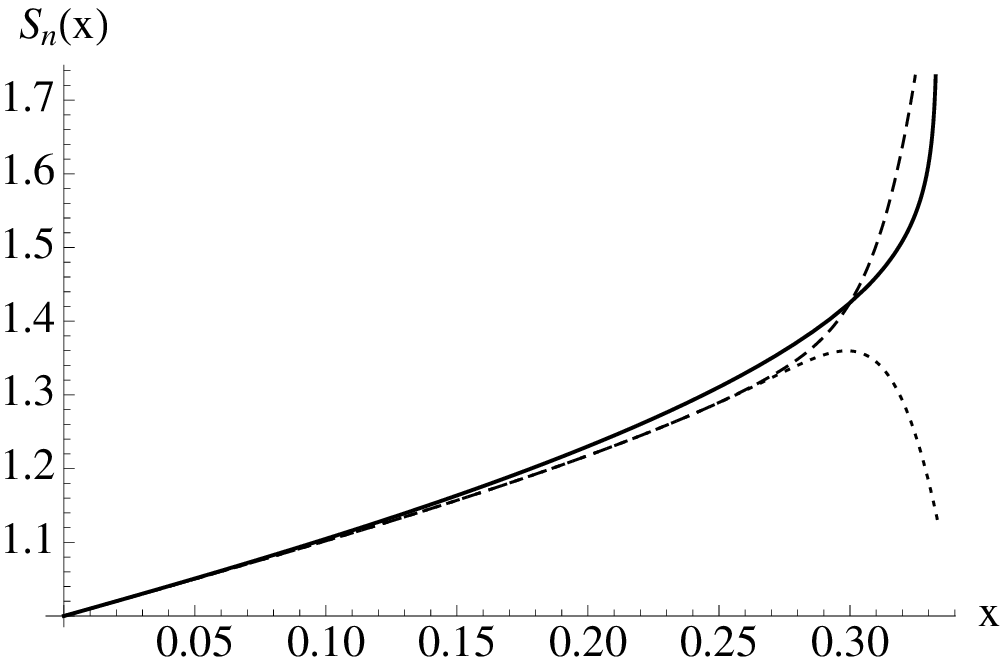}&&
\includegraphics[width=0.4\textwidth]{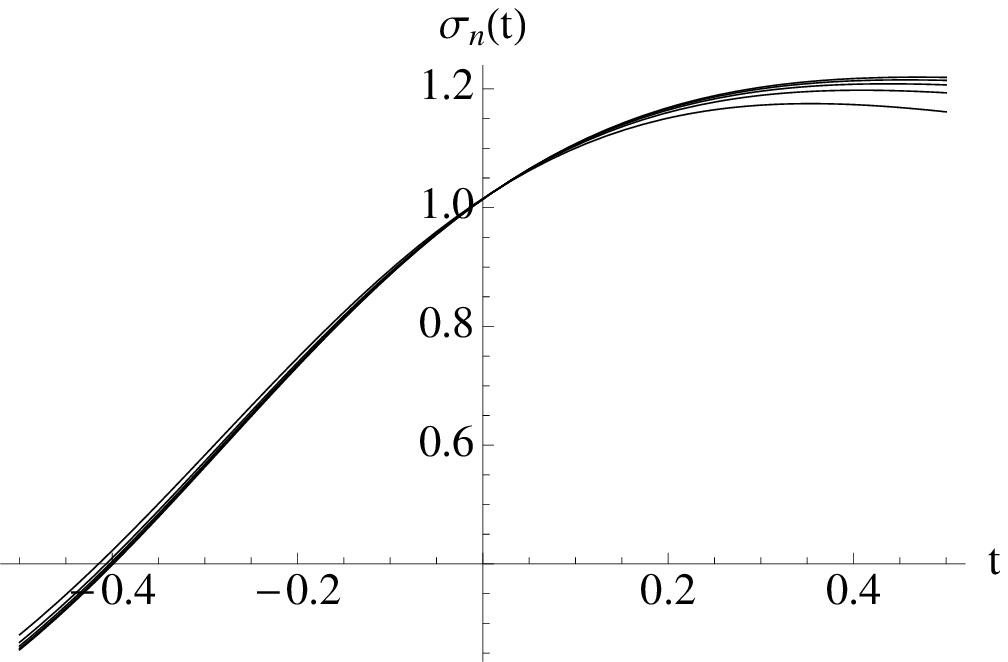}\\
\small{(a)} && \small{(b)}
\end{tabular}
\caption{(a) $S_n(x)$ for $n=50$ (dotted line) and $n=-50$ (dashed line) together with the generating function $S_\infty(x)$. (b) Near the critical point $x=1/3$ the rescaled functions $\sigma_n(t)$ collapse for large $n$: here we show $n=10,20,\dots,50$. A similar behavior can be observed for negative $n$.}
\label{fig:spoly}
\end{figure}

An illustration of this function compared to $S_n(x)$ with $n=\pm 50$ is given in figure \ref{fig:spoly}(a). 
The recurrence relation allows us to compute the first corrections to the scale-free form as well.
We have
$S_n(x)/S_\infty(x)=1+b_nx^{n+1}+c_nx^{n+2}+\dots$ Upon inserting this into
the equation for $S_n(x)$, and expanding around $x=0$ we find the
recursion relations $(n+1)b_n =2(2n+1)b_{n-1},\,b_0=-1$ and
$(n+2)(2n+3)c_n-2(2n+1)^2c_{n-1}-4b_n=0,\, d_0=1/2$ which are readily
solved:
\begin{equation}
  b_n = -\frac{1}{2}\left(
  \begin{array}{c}
  2n+2\\
  n+1
  \end{array}
  \right),\quad 
  c_n = \frac{1}{n+2}\left(
  \begin{array}{c}
  2n+2\\
  n+2
  \end{array}
  \right).
\end{equation}
Hence the first correction is a binomial -- a quite common feature that we observed for other quantities as well.


The preceding procedure relied on an expansion around the trivially
solvable point $x=0$. As we did with the order parameters, we can also
understand the scaling around the critical point $x=1/3$.  From the
asymptotic formula (\ref{Sinf}), we see that $S_\infty(x)$ displays a
power-law singularity with exponent $-1/24$. Conversely, for finite $n$ we find at this point that
\begin{equation}
  \frac{S_{n+1}(1/3)}{S_n(1/3)}
=\frac{2^{4(n+1)}s_{n+1}(1/9)s_n(1)}{3^{2n+5/2}s_n(1/9)s_{n+1}(1)} 
= \frac{2^{4n+3}}{3^{3n+5/2}}\frac{n!(3n+2)!}{((2n+1)!)^2}
\end{equation}
The value $s_n(1/9)$ necessary here can directly be inferred from the
Hirota equations by setting $z=1/9$. Upon application of Stirling's
formula for large $n$, it is not difficult to derive from this
expression the asymptotic behavior $S_n(x=1/3)\sim n^{1/36}$ for large
positive $n$. For large negative $n$ the scaling is identical because
of $S_{-n-1}(1/3)=\sqrt{3}S_n(1/3)$.  Thus to obtain something finite
in the $n\to\infty$ we obtain again the result from section
\ref{sec:scaling} that near the critical point, the deformation
parameter (here $x-1/3$) must be rescaled by a power of $n^{2/3}$. In the context of differential equations, such a relation between
properties in the two limits is known as a ``connection formula''. The rescaled functions $\sigma_n(t)= n^{-1/36}S_n(1/3-n^{-2/3}t)$ are illustrated in figure \ref{fig:spoly}(b). As before, we observe a remarkable collapse for large $n$, this time however without subtraction of any infinite-volume term like $n^{-1/36}S_\infty(1/3-n^{-2/3}t)$. It would be very interesting to find the scaling functions $\sigma_\pm(t)= \lim_{n\to\pm\infty}\sigma_n(t)$. A detailed analysis of the Hirota equations \cite{fendley:2bp} yields at least the first two Taylor coefficients
\begin{equation}
  \sigma_\pm(t) = C_\pm\left(1\pm\frac{3\Gamma(2/3)}{4\Gamma(4/3)}t+O(t^2)\right),
\end{equation}
where $C_+={G(3/2)^2}/({G(4/3)G(5/3)})=1.013993\dots$ and $C_-=\sqrt{3}C_+$.
Here $G(z)$ denotes the Barnes $G$-function, defined through the functional equation $G(z+1)=\Gamma(z)G(z)$.

\section{Conclusion}
\label{sec:concl}
We have studied an off-critical staggered model for
supersymmetric lattice fermions with hard-core exclusion, and found
some striking properties of its ground state. We showed how the
supersymmetry allows the entire ground state to be characterized by a
single polynomial, and related this
to solutions of the Hirota equations, which earlier arose 
in an XYZ chain in the same universality class. We saw how a variety
of special properties allowed the determination of exact expressions
for the order parameters and critical exponents.

The precise connection between the supersymmetric fermion model and
the XYZ chain is surprising for several reasons. First, it is known
that the critical point of the fermion chain can be transformed to the
XXZ chain at $\Delta=-1/2$ with twisted boundary conditions and an
\textit{even} number of sites \cite{fendley:03,yang:04}. Even though
there is a relation between the cases $L=2n$ with twisted boundary
conditions and $L=2n+1$ with periodic boundary conditions for this XXZ
chain \cite{difrancesco:06}, what happens off criticality remains
mysterious. In particular, no XYZ chain with twisted boundary
conditions that remains integrable is known, since there it has no
obvious $U(1)$ conservation law. Note, however, the fermion chain does
possess a $U(1)$ fermion number conservation law.  Another distinction
between the two is in the dualities.  For the XYZ chain, the duality
transformation (\ref{eqn:homographic}) is related to simple unitary
transformations on the Hilbert space (spin rotations) and
obvious invariance properties of the corresponding Hamiltonian. In
the fermion case however, the transformation on the coupling constant
$y$ is highly non-linear, and a realization of the duality on the Hilbert
space through a unitary transformation seems to be less evident.

These observations give a strong hint that there remains much
structure to be uncovered in both the fermion and XYZ chains. As the latter are related to the eight-vertex model at the special crossing parameter $\eta=\pi/3$, powerful theoretical tools such as Baxter's famous TQ-equation can be used for their analysis. For example, it allows the exact evaluation of the nearest-neighbor spin correlators \textit{in finite size} \cite{fendley:2bp}. In the case of the fermion chain, similar relations remain to be discovered.

\section*{Acknowledgments}
 This work was supported by the NSF grant DMR/MSPA-0704666. The research of CH was supported in part by the NSF grant PHY05-51164.

\appendix
\section{Combinatorial properties at the critical point}

Given the connection between the critical model and the XXZ chain at
$\Delta=-1/2$, it is natural to expect that the values $m^\pm_n(y=1)$
are related to combinatorial numbers enumerating alternating sign
matrices and related objects. By explicitly finding the ground states
at $y=1$, some of these combinatorial properties were explored in
\cite{Beccaria}. Here we analyze the particular case
of $N=3n$ and periodic boundary conditions in more depth.

First, we notice that at $y=1$ we find
  \begin{equation*}
  m_{-2k}(1)=3m_{2k}(1)\quad \mbox{and} \quad m_{1-2k}(1) = m_{2k-1}(1),
\end{equation*}
for $k=1,2,\dots$. Therefore it will be sufficient to concentrate on
the parity-even sequence $m_n(1)$. For odd $n=2k-1$ we thus
find $m_{2k-1}(1)=1,5,198,63206,163170556,\dots$ This sequence
corresponds to the numbers $A_{UU}(4k)$ of $4k\times 4k$ alternating
sign matrices of type UU \cite{kuperberg:02}. It is known to factorize
into
\begin{equation}
  m_{2k-1}(1)=A_{UU}(4k)=A_V(2k+1)\times A_{UU}(4k;1,1,1)
  \label{eqn:oddcv}
\end{equation}
where
\begin{eqnarray*}
  A_V(2k+1)&=\frac{1}{2^k} \prod_{j=1}^k \frac{(6j-2)!(2j-1)!}{(4j-1)!(4j-2)!},
\\
  A_{UU}(4k;1,1,1) &=(-3)^{k^2}2^{2k}\prod_{i=1}^{2k}\prod_{{j=1,2\mid j}}^{2k+1}\frac{3j - 3i + 2}{j - i +2n +1}.
\end{eqnarray*}
Here $A_V(2k+1)$ is the number of alternating sign $(2k+1)\times (2k+1)$ matrices symmetric about the vertical axis, also the number of $(2k)\times (2k)$ off-diagonally symmetric alternating sign matrices, and $A_{UU}(4k;1,1,1)$ expansion coefficients of a generating function appearing in \cite{kuperberg:02,robbins:00}.

For even $n=2k$ we find the values $m_{2k}(1)=1,14,1573,1427660,\dots$ As is the case of an odd number of particles, this sequence can be factorized according to 
\begin{equation}
  m_{2k}(y=1) = N_8(k)\times\frac{A(2k-1)}{A_V(2k-1)},\quad k=1,2,\dots
  \label{eqn:evencv}
\end{equation}
with the numbers of cyclically symmetric transpose complementary plane partitions $N_8(k)$ given by \cite{degier:02}
\begin{equation*}
  N_8(k) = \prod_{j=0}^{k-1}(3j+1)\frac{(6j)!(2j)!}{(4j)!(4j+1)!},
\end{equation*}
and the well-known numbers of $n\times n$ alternating sign matrices
\begin{equation*}
   A(k) = \prod_{j=0}^{k-1}\frac{(3j+1)!}{(k+j)!}.
\end{equation*}

We emphasized the factorizations because it turns out that the
subfactors of $m_{2k-1}(y^2)$ and $m_{-2k}(y^2)$ are in one-to-one
correspondence with the factors in \eqref{eqn:oddcv} and
\eqref{eqn:evencv} when setting $y=1$.



\section*{References}

\begin{thebibliography}{10}

\bibitem{Baxbook} R.J.\ Baxter,
\newblock
{\em Exactly Solved Models in Statistical Mechanics} (Academic,
London, 1982)

\bibitem{Witten}    
E. Witten, 
\newblock {\em Constraints on supersymmetry breaking},
\newblock Nucl. Phys. B \textbf{202} (1982) 253


\bibitem{fendley:02}
P.~{Fendley}, K.~{Schoutens}  and J.~{de Boer},
\newblock {\em {Lattice Models with N=2 Supersymmetry}},
\newblock Phys. Rev. Lett. {\textbf{90}} (2003)   120402,
\newblock arXiv:hep-th/0210161.

\bibitem{fendley:03}
P.~{Fendley}, B.~{Nienhuis}  and K.~{Schoutens},
\newblock {\em {Lattice fermion models with supersymmetry}},
\newblock J.\ Phys.\ A: Math.\ Gen.\ {\textbf{36}} (2003)   12399,
\newblock arXiv:cond-mat/0307338.


\bibitem{fendley:10}
P.~{Fendley} and C.~{Hagendorf},
\newblock {\em {Exact and simple results for the XYZ and strongly interacting
  fermion chains}},
\newblock J.\ Phys.\ A: Math.\ Theor.\ {\textbf{43}} (2010)   402004



\bibitem{razumov:01}
A.~V.\ {Razumov} and Y.~G.\ {Stroganov},
\newblock {\em {Spin chains and combinatorics: twisted boundary conditions}},
\newblock J.\ Phys.\ A: Math.\ Gen.\ {\textbf{34}} (2001)   5335,
\newblock arXiv:cond-mat/0102247.

\bibitem{stroganov:01_2}
Y.~Stroganov,
\newblock {\em The 8-vertex model with a special value of the crossing
  parameter and the related xyz chain},
\newblock in {\em {Integrable structures of exactly solvable two-dimensional
  models of quantum field theory (Kiev 2000)}}, {\em {\em Volume}~35} of {\em
  NATO Sci.\ Ser.\ II Math.\ Phys.\ Chem.}, pages 315--319, Kluwer Acad.\ Publ.,
  Dordrecht, 2001.



\bibitem{degier:02}
J.~{de Gier}, M.~T.\ {Batchelor}, B.~{Nienhuis}  and S.~{Mitra},
\newblock {\em {The XXZ spin chain at {$\Delta=-1/2$}: Bethe roots, symmetric
  functions, and determinants}},
\newblock J.\ Math.\ Phys.\ {\textbf{43}} (2002)   4135,
\newblock arXiv:math-ph/0110011.




\bibitem{yang:04}
X.~Yang and P.~Fendley,
\newblock{\em Non-local space-time supersymmetry on the lattice},
\newblock J.\ Phys.\ A {\textbf 37} (2004) 8937,
\newblock arXiv:cond-mat/0404682.



\bibitem{bazhanov:05}
V.~V.\ {Bazhanov} and V.~V.\ {Mangazeev},
\newblock {\em {Eight-vertex model and non-stationary Lam{\'e} equation}},
\newblock J.\ Phys.\ A: Math.\ Gen.\ {\textbf{38}} (2005)   L145,

\bibitem{bazhanov:06}
V.~V.\ {Bazhanov} and V.~V.\ {Mangazeev},
\newblock {\em {The eight-vertex model and Painlev{\'e} VI}},
\newblock J.\ Phys.\ A: Math.\ Gen.\ {\textbf{39}} (2006)   12235,
\newblock arXiv:hep-th/0602122.

\bibitem{mangazeev:10}
V.~V.\ {Mangazeev} and V.~V.\ {Bazhanov},
\newblock {\em {The eight-vertex model and Painlev{\'e} VI equation II:
  eigenvector results}},
\newblock J.\ Phys.\ A: Math.\ Theor.\ {\textbf{43}} (2010)   085206,
\newblock arXiv:0912.2163.

\bibitem{razumov:10}
A.~V.\ {Razumov} and Y.~G.\ {Stroganov},
\newblock {\em {A possible combinatorial point for the XYZ spin chain}},
\newblock Theor.\ Math.\ Phys.\ {\textbf{164}} (2010)   977,
\newblock 0911.5030.

\bibitem{Huijse} L.~Huijse, University of Amsterdam thesis (2010)

\bibitem{FI}
P.~Fendley and K.~Intriligator, 
\newblock {\em Scattering and Thermodynamics of
Fractionally-Charged Supersymmetric Solitons}, 
\newblock Nucl.~Phys.~B {\bf 372}, 533 (1992),
\newblock arXiv:hep-th/9111014 




\bibitem{Beccaria} M.~Beccaria and G.F.~De Angelis,
\newblock{\em Exact Ground State and Finite Size Scaling in a Supersymmetric
Lattice Model,''}
\newblock
Phys.\ Rev.\ Lett.\  {\bf 94}, 100401 (2005),
\newblock arXiv:cond-mat/0407752.


\bibitem{baxtersolving}
R.J.~Baxter,
\newblock {\em Solving models in statistical mechanics},
\newblock Adv.\ Stud.\ Pure Math.\ {\textbf{19}} (1989)   95,



\bibitem{okamoto:87}
K.~Okamoto,
\newblock {\em {Studies on the Painlev{\'e} equations.\ I: Sixth Painlev{\'e}
  equation PVI}},
\newblock Annali di Matematica pura ed applicata {\textbf{146}} (1987)
  337.


\bibitem{mangazeev:10_2}
V.~V.\ {Mangazeev},
\newblock {Picard solution of Painlev\'e VI and related
  tau-functions},
\newblock arXiv:1002.2327 2010.


\bibitem{difrancesco:06}
P.~{Di Francesco}, P.~{Zinn-Justin}  and {J.-B.} {Zuber},
\newblock {\em {Sum rules for the ground states of the O(1) loop model on a
  cylinder and the XXZ spin chain}},
\newblock J.\ Stat.\ Mech.\ {\textbf{8}} (2006)  ~11,
\newblock arXiv:math-ph/0603009.


\bibitem{kuperberg:02}
G.~ Kuperberg,
\newblock {\em {Symmetry Classes of Alternating-Sign Matrices under One Roof}},
\newblock Ann.\ Math.\ {\textbf{156}} (2002)   835,
\newblock arXiv:math/0008184


\bibitem{robbins:00}
D.~P.\ {Robbins},
\newblock {Symmetry Classes of Alternating Sign Matrices},
\newblock arXiv:math.CO/0008045.






\bibitem{fendley:2bp}
C.\ Hagendorf,
\newblock in preparation (2010).

\end{thebibliography}
\end{document}